\documentclass[aip,jmp,amsmath,amssymb,reprint]{revtex4-1}
\pdfoutput=1          % PDFLatex [arXiv]
\usepackage{graphics} % Include figure files
\usepackage{graphicx} % Include figure files
\usepackage{dcolumn}  % Align table columns on decimal point
\usepackage{bm}       % bold math
\usepackage{natbib}

\begin{document}

\title{Band-structure and electronic transport calculations in cylindrical wires : the issue of bound states in transfer-matrix calculations}

\author{Alexandre Mayer}
\email{alexandre.mayer@unamur.be}
\affiliation{Department of Physics, University of Namur, Rue de Bruxelles 61, 5000 Namur, Belgium}

\date{16 July 2019}

\begin{abstract}
The transfer-matrix methodology is used to solve linear systems of
differential equations, such as those that arise when solving
Schr\"odinger's equation, in situations where the solutions of interest
are in the continuous part of the energy spectrum. The technique is
actually a generalization in three dimensions of methods used to obtain
scattering solutions in one dimension. Using the layer-addition
algorithm allows one to control the stability of the computation
and to describe efficiently periodic repetitions of a basic unit.
This paper, which is an update of an article originally published
in Physical and Chemical News 16, 46-53 (2004), provides a pedagogical
presentation of this technique. It describes in details how the band
structure associated with an infinite periodic medium can be extracted
from the transfer matrices that characterize a single basic unit. The
method is applied to the calculation of the transmission and band
structure of electrons subject to cosine potentials in a cylindrical
wire. The simulations show that bound states must be considered because
of their impact as sharp resonances in the transmission probabilities
and to remove unphysical discontinuities in the band structure.
Additional states only improve the completeness of the representation.
\end{abstract}


\maketitle

\section{Introduction}

The transfer-matrix methodology is one of the techniques used to solve
linear systems of differential equations, such as those that arise when solving
Schr\"odinger's equation, in situations where the
solutions of interest are in the continuous part of the energy spectrum.
For this numerical scheme to be relevant,
the physical system considered should be located between two separate
boundaries (standing for the regions of incidence and transmission).
Given a set of basis states used for the expansion of the wave
function, the transfer matrices provide, for each state incident on one
boundary of the system, the coefficients of the corresponding
reflected and transmitted states.

The advantage of this technique is that it does not require the storage of
the wave function in the intermediate part of the system (where
solutions are only propagated through). Its storage space requirements
therefore depend essentially on the number $N$ of basis
states used for the expansion of the solutions (more precisely on
$N^{3}$), and not directly on the dimensions of the system.
This technique was first developed by Pendry\cite{Pendry1,Pendry2,Pendry3}
for Low Energy Electron Diffraction simulations.
It was used and developed by other authors,\cite{Vigneron1,Sheng1,Russel1,Ward1,Bayman1,Wu1,Yalabik1,Anemogiannis1,Price2,Price3,Price1}
including Mayer et al.\cite{Mayer4,Mayer9,Mayer12}
for the simulation of the Fresnel projection microscope,\cite{Binh4,Mayer1,Mayer2,Mayer5}
for the modeling of field electronic emission,\cite{Mayer70,Mayer72,Mayer75,Mayer100}
for the modeling of photon-stimulated field emission\cite{Mayer20,Mayer30} and finally
for the modeling of optical rectification by geometrically asymmetric
metal-vacuum-metal junctions.\cite{Mayer56,Mayer59,Mayer76,Mayer78}

An interesting feature of the method is that it can easily handle
periodic repetitions of a basic unit. From the transfer matrices
associated with a single unit of the structure, it is indeed
straightforward (using the layer-addition
algorithm\cite{Pendry1,Pendry2}) to derive those
corresponding to an arbitrary number of units. The band structure that
characterizes the infinite repetition of these units can also be
extracted from the transfer matrices.

It is the objective of this paper to provide a pedagogical
presentation of the transfer-matrix methodology and to
describe how band structures can be derived in this approach.
The theoretical aspects of this scheme are developed in Sec. II.
The technique is then applied in
Sec. III to the study of electrons that are confined in a cylindrical wire
and subject to cosine potentials. The simulations show how fast band
structure effects appear with the number of periods.
The features of the transmission diagram are related to those of the
band structure and interpreted in terms of quantum
conductance and band-gap effects. The issue of bound states is also considered.
It is found that they need to be considered in order to reproduce sharp
resonances in the transmission probabilities and to remove unphysical 
discontinuities in the band structure. Additional states only improve the
completeness of the representation.

\section{Theory}

Let us consider three regions: Region I ($z\leq 0$), Region II
($0\leq z\leq D$) and Region III ($z\geq D$). We consider the
scattering strengths to be in the intermediate Region II and we want to
compute how electronic states incident on one side of Region II
are scattered towards the other side.

\begin{figure}[ht]
 \begin{center}
  \begin{tabular}{c}
   \includegraphics[width=8cm]{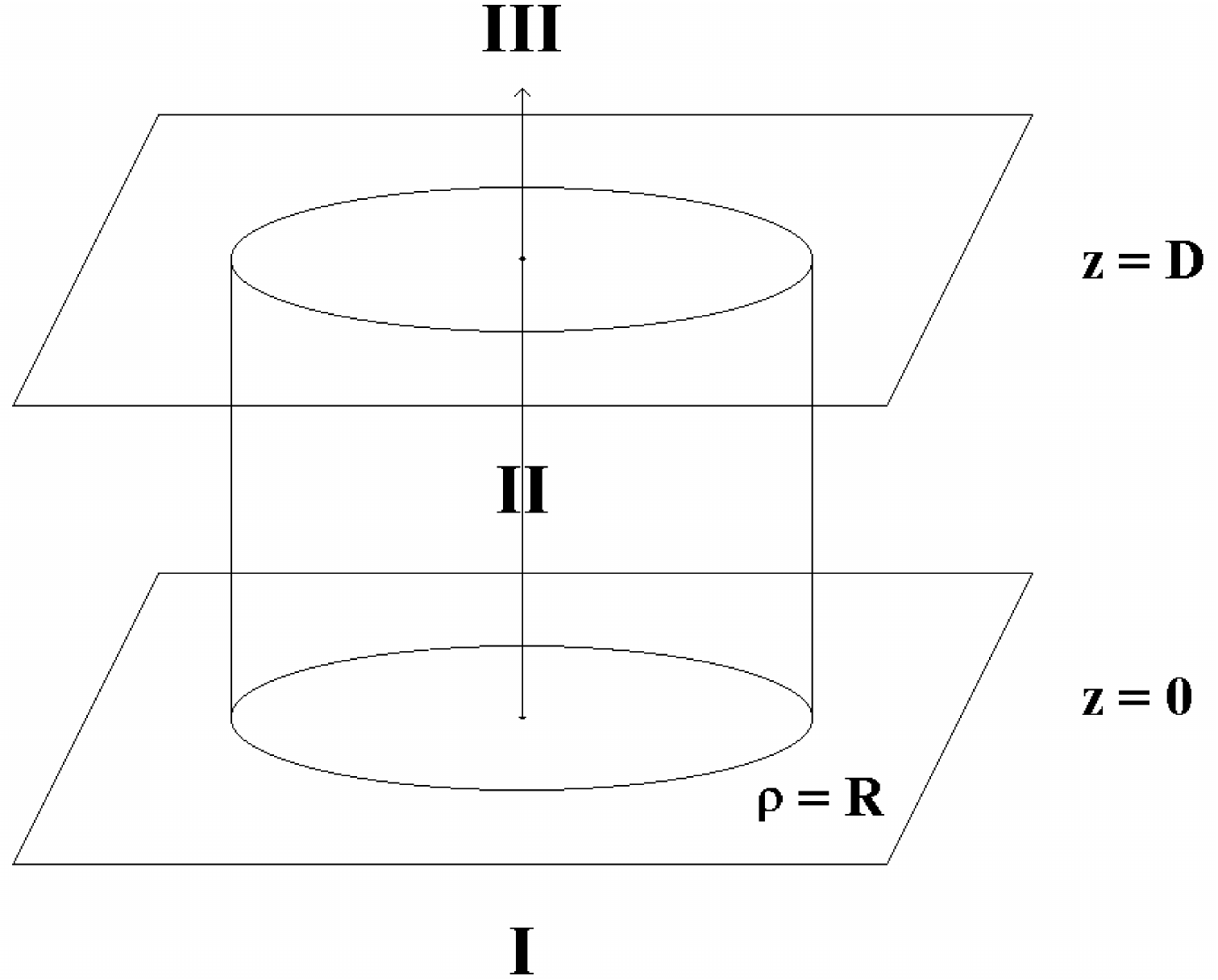}
  \end{tabular}
 \end{center}
 \caption{\label{figure1}Schematic representation of the situation considered. Region I
  and Region III are the regions of incidence and transmission. The intermediate
  Region II contains a cylindrical wire with cosine potentials.}
\end{figure}

Let us consider two sets of basis states in the boundary regions,
namely $\{ \Psi_{j}^{\rm I,\pm} \}$ in Region I and
$\{ \Psi_{j}^{\rm III,\pm} \}$ in Region III. These states are used to
expand the wave function in Region I and Region III, for a given value
of the energy $E$. The subscript $j$ enumerates the allowed
values of $\left\{ k_{\rho},m \right\}$ in cylindrical coordinates
or $\left\{ k_{\rm x},k_{\rm y} \right\}$ in cartesian coordinates, considering
applicable boundary conditions and the energy $E$. The $\pm$ signs refer to the
propagation direction relative to the $z$ axis, which is oriented from Region I to Region III
(see Fig. \ref{figure1}).

We assume our boundary states to be separable in the following way:
\begin{eqnarray}
 \label{basis1}
 \Psi_{j}^{\rm I,\pm}(\rho,\phi,z)
   &=& \psi_{j}(\rho,\phi) \exp(\pm i k_{z,j} z), \\
 \label{basis2}
 \Psi_{j}^{\rm III,\pm}(\rho,\phi,z)
   &=& \psi_{j}(\rho,\phi) \exp(\pm i k_{z,j} z),
\end{eqnarray}
where $\psi_{j}(\rho,\phi)$ refers to analytical basis
functions that account for the $(\rho,\phi)$-dependence of
the wave functions (see Sec. \ref{Section_Application}
for specific expressions in cylindrical coordinates). These relations
will be used to derive band structures from transfer-matrix calculations;
they are actually only required for this specific application.

We will describe in this section how scattering solutions that
correspond to single incident states $\Psi_{j}^{\rm I,+}$ in Region
I or $\Psi_{j}^{\rm III,-}$ in Region III can be derived. We will
describe shortly the layer-addition algorithm and finally explain how
the band structure associated with the infinite repetition of a basic
unit can be extracted from these solutions.

\subsection{Basic formulation of the transfer-matrix technique}

The first step of the technique consists in establishing solutions associated
with single outgoing states $\Psi_{j}^{\rm III,+}$ or incoming states $\Psi_{j}^{\rm III,-}$
in Region III. Since the wave function and its derivatives are entirely defined
in Region III, one can propagate these states numerically from
$z=D$ to $z=0$, where the solutions are expanded in terms of incident states $\Psi_{j}^{\rm I,+}$
and reflected states $\Psi_{j}^{\rm I,-}$.
The numerical techniques that enable the propagation of the wave functions through Region II
can be found in Refs \citenum{Mayer1}, \citenum{Mayer70} and \citenum{Mayer100}.
The expansion coefficients of these solutions are stored in ${\bf T}^{\pm\pm}$ matrices
and we end up with the following set of solutions:
\begin{eqnarray}
 \label{psi+}
 \overline{\Psi}^{+}_{j}
  &\stackrel{z\leq 0}{=}&
   \sum_{i} T_{i,j}^{++} \Psi^{\rm I,+}_{i}
  +\sum_{i} T_{i,j}^{-+} \Psi^{\rm I,-}_{i}
  \stackrel{z\geq D}{=}
   \Psi^{\rm III,+}_{j}, \\
 \label{psi-}
 \overline{\Psi}^{-}_{j}
  &\stackrel{z\leq 0}{=}&
   \sum_{i} T_{i,j}^{+-} \Psi^{\rm I,+}_{i}
  +\sum_{i} T_{i,j}^{--} \Psi^{\rm I,-}_{i}
  \stackrel{z\geq D}{=}
   \Psi^{\rm III,-}_{j}.
\end{eqnarray}

In the second step of the procedure, these solutions are combined linearly in order to
derive new solutions satisfying the scattering boundary conditions,
namely solutions associated with either a single incident state
$\Psi_{j}^{\rm I,+}$ in Region I or a single incident state
$\Psi_{j}^{\rm III,-}$ in Region III (with this time reflected
states in the region of incidence and transmitted states in the
other region).
Formally, these solutions are expressed in terms of scattering
${\bf S}^{\pm\pm}$ matrices in the following way:
\begin{eqnarray}
\label{t++}
\Psi^{+}_{j}
 &\stackrel{z\leq 0}{=}&
  \Psi^{\rm I,+}_{j} + \sum_{i} S^{-+}_{i,j} \Psi^{\rm I,-}_{i}
  \stackrel{z\geq D}{=}
  \sum_{i} S^{++}_{i,j} \Psi^{\rm III,+}_{i}, \\
\label{t--}
\Psi^{-}_{j}
 &\stackrel{z\leq 0}{=}&
  \sum_{i} S^{--}_{i,j} \Psi^{\rm I,-}_{i}
  \stackrel{z\geq D}{=}
  \Psi_{j}^{\rm III,-} + \sum_{i} S^{+-}_{i,j} \Psi^{\rm III,+}_{i}.
\end{eqnarray}
The ${\bf S}^{\pm\pm}$ matrices, which contain the expansion
coefficients of these scattering solutions are related to the
${\bf T}^{\pm\pm}$ matrices of Eqs \ref{psi+} and \ref{psi-} by
${\bf S}^{++}={{\bf T}^{++}}^{-1}$,
${\bf S}^{-+}={\bf T}^{-+}{{\bf T}^{++}}^{-1}$,
${\bf S}^{--}={\bf T}^{--}-{\bf T}^{-+}{{\bf T}^{++}}^{-1}{\bf T}^{+-}$ and
${\bf S}^{+-}=-{{\bf T}^{++}}^{-1}{\bf T}^{+-}$.

\subsection{\label{layer-addition}The layer-addition algorithm for the control of accuracy
  and the description of periodic systems}

To control the numerical instabilities that appear with large distances
$D$ (when inverting ${\bf T}^{++}$ to obtain the ${\bf S}^{\pm\pm}$ matrices)
or to treat efficiently periodic systems, it is useful to use the
layer-addition algorithm.\cite{Pendry1,Pendry2}
Given a subdivision $ 0=z_{0} < z_{1} < z_{2} < \cdots < z_{n-1} < z_{n} = D$
of the interval $[0,D]$ and referring by
${\bf S}^{++}_{z_{i},z_{j}}$, ${\bf S}^{-+}_{z_{i},z_{j}}$,
${\bf S}^{--}_{z_{i},z_{j}}$ and ${\bf S}^{+-}_{z_{i},z_{j}}$ to the $S$
matrices associated with the interval
$[ z_{i},  z_{j} ]$, one can derive those associated
with the entire interval $[0,D]$ from the recursive application
of the following relations:
\begin{eqnarray}
{\bf S}^{++}_{z_{0},z_{i}}
&=& {\bf S}^{++}_{z_{i-1},z_{i}}
 \left[ {\bf I}- {\bf S}^{+-}_{z_{0},z_{i-1}} {\bf S}^{-+}_{z_{i-1},z_{i}} \right]^{-1}
  {\bf S}^{++}_{z_{0},z_{i-1}}, \\
{\bf S}^{--}_{z_{0},z_{i}}
&=& {\bf S}^{--}_{z_{0},z_{i-1}}
    \left[
    {\bf I}- {\bf S}^{-+}_{z_{i-1},z_{i}} {\bf S}^{+-}_{z_{0},z_{i-1}}
    \right]^{-1}
    {\bf S}^{--}_{z_{i-1},z_{i}}, \\
{\bf S}^{-+}_{z_{0},z_{i}}
&=&  {\bf S}^{-+}_{z_{0},z_{i-1}}
 +  {\bf S}^{--}_{z_{0},z_{i-1}}
  {\bf S}^{-+}_{z_{i-1},z_{i}}
 \left[
  {\bf I}-{\bf S}^{+-}_{z_{0},z_{i-1}} {\bf S}^{-+}_{z_{i-1},z_{i}}
 \right]^{-1}
 {\bf S}^{++}_{z_{0},z_{i-1}}, \\
{\bf S}^{+-}_{z_{0},z_{i}}
&=&{\bf S}^{+-}_{z_{i-1},z_{i}}
 + {\bf S}^{++}_{z_{i-1},z_{i}}
   {\bf S}^{+-}_{z_{0},z_{i-1}}
   \left[
   {\bf I} - {\bf S}^{-+}_{z_{i-1},z_{i}} {\bf S}^{+-}_{z_{0},z_{i-1}}
   \right]^{-1}
   {\bf S}^{--}_{z_{i-1},z_{i}}.
\end{eqnarray}

These relations enable a straightforward derivation of the ${\bf S}^{\pm\pm}$
matrices associated with the periodic repetition of an arbitrarily large number
of units once the transmission through a single unit has been
established. Even in the case of non-periodic systems, it
is generally useful to use this algorithm
since the relative error on the transfer-matrix calculations increases
exponentially with the distance $D$ if it is considered in a single
step. The number of subdivisions to consider in order to achieve a
given accuracy is given, with other considerations on the stability of
transfer-matrix calculations, in Ref. \citenum{Mayer4}.

For the derivation of band structures in the next subsection, we will use the
${\bf T}^{\pm\pm}$ matrices. These matrices keep stable when considering large distances $D$
(only the inversion of ${\bf T}^{++}$ is unstable when $D$ is too large). When
these matrices are obtained for subdivisions of the $[0,D]$ interval, they can
be updated according to the following formula:
\begin{eqnarray}
\begin{pmatrix}
{\bf T}^{++}_{z_{0},z_{i}} & {\bf T}^{+-}_{z_{0},z_{i}} \cr
{\bf T}^{-+}_{z_{0},z_{i}} & {\bf T}^{--}_{z_{0},z_{i}} \cr
\end{pmatrix}
=
\begin{pmatrix}
{\bf T}^{++}_{z_{0},z_{i-1}} & {\bf T}^{+-}_{z_{0},z_{i-1}} \cr
{\bf T}^{-+}_{z_{0},z_{i-1}} & {\bf T}^{--}_{z_{0},z_{i-1}} \cr
\end{pmatrix}
\begin{pmatrix}
{\bf T}^{++}_{z_{i-1},z_{i}} & {\bf T}^{+-}_{z_{i-1},z_{i}} \cr
{\bf T}^{-+}_{z_{i-1},z_{i}} & {\bf T}^{--}_{z_{i-1},z_{i}} \cr
\end{pmatrix}.
\end{eqnarray}

\subsection{Derivation of band structures from transfer matrices}

Let us now consider a basic unit, of length $a$ in the $z$ direction.
One can compute the transfer matrices associated with this structure,
for given values of the energy $E$. Our objective is to extract from
these matrices the band structure characterizing the
infinite, periodic repetition of this unit.

For this purpose let us first reconsider the solutions of
Eqs \ref{psi+} and \ref{psi-}, which are recast in the following way:
\begin {eqnarray}
\label{scatt2}
( \overline{\Psi}_{j}^{+} \ldots \overline{\Psi}_{j}^{-} )
\stackrel{z\leq 0}{=}
( \Psi_{j}^{\rm I,+} \ldots \Psi_{j}^{\rm I,-} )
\begin{pmatrix}
{\bf T}^{++} & {\bf T}^{+-} \cr
{\bf T}^{-+} & {\bf T}^{--} \cr
\end{pmatrix}
\stackrel{z \geq a}{=}
( \Psi_{j}^{\rm III,+} \ldots \Psi_{j}^{\rm III,-} )
.
\end{eqnarray}

We want to find combinations
$( \overline{\Psi}_{j}^{+} \ldots \overline{\Psi}_{j}^{-} ) {\bf x}$
of these solutions that satisfy the relation:
\begin{eqnarray}
(\overline{\Psi}_{j}^{+} \ldots \overline{\Psi}_{j}^{-} )
{\bf x} |_{z=a}
=
 e^{ik_{z}a}\
(\overline{\Psi}_{j}^{+} \ldots \overline{\Psi}_{j}^{-} )
{\bf x} |_{z=0},\label{eigensystem_explicit}
\end{eqnarray}
where ${\bf x}$ is a vector that contains the coefficients of these
combinations. These combinations describe particular states that
keep unchanged after propagation through one period of the system
except for a phase factor $\lambda=\exp ( i k_{z} a )$. These states are
therefore Bloch states associated with a wave vector $k_{z}$ in
the first Brillouin zone $[ -\pi/a, \pi/a ]$ of the periodic system,
for the energy $E$ considered. The couples of all possible points
$(k_{z},E)$ will represent the band structure of the system.

In order to establish a matricial equation for the calculation
of a complete set of Bloch-state solutions, we will
write Eq. \ref{eigensystem_explicit} like
\begin{eqnarray}
(\overline{\Psi}_{j}^{+} \ldots \overline{\Psi}_{j}^{-} )
{\bf X} |_{z=a}
=
(\overline{\Psi}_{j}^{+} \ldots \overline{\Psi}_{j}^{-} )
{\bf X} |_{z=0}\ {\bf \Lambda},\label{bandstruct}
\end{eqnarray}
where ${\bf \Lambda}$ is a diagonal matrix containing elements of the form
$\lambda = \exp ( i k_{z} a )$ and ${\bf X}$ is a matrix whose columns
contain the coefficients ${\bf x}$ of each Bloch-state solution.

We have from Eq. \ref{scatt2} that
\begin{eqnarray}
 (\overline{\Psi}_{j}^{+} \ldots \overline{\Psi}_{j}^{-} ) {\bf X} |_{z=a}
&=&
 (\Psi_{j}^{\rm III,+} \ldots \Psi_{j}^{\rm III,-} )|_{z=a}\ {\bf X}, \label{Bloch_State_1}\\
%%%
 (\overline{\Psi}_{j}^{+} \ldots \overline{\Psi}_{j}^{-} ) {\bf X} |_{z=0}
&=&
( \Psi_{j}^{\rm I,+} \ldots \Psi_{j}^{\rm I,-} )|_{z=0}\
\begin{pmatrix}
{\bf T}^{++} & {\bf T}^{+-} \cr
{\bf T}^{-+} & {\bf T}^{--} \cr
\end{pmatrix}
\ {\bf X}.\label{Bloch_State_2}
\end{eqnarray}

If we remember the expression of the basis states $\Psi_{j}^{\rm I,\pm}$ and $\Psi_{j}^{\rm III,\pm}$
(see Eqs \ref{basis1} and \ref{basis2}), we can
actually relate them by
\begin{eqnarray}
( \Psi_{j}^{\rm I,+} \ldots \Psi_{j}^{\rm I,-} )|_{z=0}
=
( \Psi_{j}^{\rm III,+} \ldots \Psi_{j}^{\rm III,-} )|_{z=a}\
{\bf diag}[e^{-i k_{z,j} a}, \ldots, e^{i k_{z,j} a}],\label{basis_transformation}
\end{eqnarray}
where ${\bf diag}[ ]$ stands for a diagonal matrix containing the
elements in brackets.

By accounting for Eqs \ref{Bloch_State_1}, \ref{Bloch_State_2} and
\ref{basis_transformation} in Eq. \ref{bandstruct}, we obtain
\begin {eqnarray}
\label{scatt3}
{\bf X}
=
{\bf diag}[e^{-i k_{z,j} a}, \ldots, e^{i k_{z,j} a}]
\begin{pmatrix}
{\bf T}^{++} & {\bf T}^{+-} \cr
{\bf T}^{-+} & {\bf T}^{--} \cr
\end{pmatrix}
{\bf X}\ {\bf \Lambda}
\end{eqnarray}
or equivalently
\begin {eqnarray}
 {\bf X}
 {\bf \Lambda}^{-1}
 {\bf X}^{-1}
=
 {\bf diag}[e^{-i k_{z,j} a}, \ldots, e^{i k_{z,j} a}]
 \begin{pmatrix}
  {\bf T}^{++} & {\bf T}^{+-} \cr
  {\bf T}^{-+} & {\bf T}^{--} \cr
 \end{pmatrix}.\label{bandstructure_system}
\end{eqnarray}
Eq. \ref{bandstructure_system} implies that the eigenvalues ${\overline \lambda}$ of the
matrix on the right-hand side of this expression will provide the wave vectors $k_{z}$
that characterize Bloch states associated with the energy $E$
[through $\lambda = {\overline \lambda}^{-1} = \exp (i k_{z} a )$].
Note that in most techniques the values of $E$ are obtained as a
function of $k_{z}$ and that the restriction of $k_{z}$ in the first Brillouin zone
$[-\pi/a,\pi/a]$ of the periodic system is automatically verified.

It has to be noted that the values of $\lambda = {\overline \lambda}^{-1} $ are not
always in the form $\exp ( i k_{z} a )$, especially in situations
involving tunneling processes or in band-gap regions. Many if not all of
them can indeed exhibit an exponential dependence $\exp ( K a )$ and
are therefore not relevant to the band structure. One distinguishes
the values $\lambda$ to consider for the representation of the
band structure by the condition $|\lambda|= 1$ (within numerical precision).

A numerically more stable technique was formulated by Pendry in Ref. \citenum{Pendry1} and
used by Mayer in Ref. \citenum{Mayer_2004} to compute the band structure of carbon nanotubes.
It consists in solving the generalized eigenvalue problem
\begin {eqnarray}
\begin{pmatrix}{\bf S}^{++} & 0 \cr -{\bf S}^{-+} & {\bf I} \cr \end{pmatrix}
{\bf x}
= \lambda \begin{pmatrix} {\bf I} & -{\bf S}^{+-} \cr 0 & {\bf
S}^{--} \cr \end{pmatrix}
{\bf diag}[e^{-i k_{z,j} a}, \ldots, e^{i k_{z,j}
a}]\ {\bf x},
\end{eqnarray}
where $\lambda$ and ${\bf x}$ are here generalized eigenvalues and
eigenvectors (see Appendix \ref{Appendix_A} for a demonstration).
The $\lambda$ values that are relevant to the band structure are
again those that satisfy $|\lambda|= 1$ (within numerical precision).
They define individual points $(k_{z},E)$ of the band structure through
$\lambda=e^{ik_{z}a}$. The restriction of $k_{z}$ in the first Brillouin zone
$[-\pi/a,\pi/a]$ of the periodic system is again automatically verified.

For a given problem, these techniques provide a particular
{\it representation} of the band structure, since its structure
in the three-dimensional reciprocal space
is projected on the $k_{z}$ axis. This is
a consequence of formulating the three-dimensional scattering of the wave function
as the one-dimensional propagation of its components.
In general this representation is appropriate in situations where a
treatment by transfer matrices is relevant.

\section{\label{Section_Application}Application: band structure and transport properties of cylindrical
  wires}

The applications considered in this paper will
focus on the scattering of electrons subject to cosine potentials in a
cylindrical wire. We will compute the transmission through a finite
number of periods and compare these results with the band structure
that characterizes the infinite medium. We will also study the
impact of bound states in the intermediate region and discuss the
necessity to consider them or not in a transfer-matrix calculation.

We assume that the radius $R$ of the wires is identical to the period
$a$ in the $z$ direction. Using cylindrical coordinates, the boundary
states we use for the
representation of the wave function in Region I and Region III are given by:
\begin{eqnarray}
 \Psi_{m,j}^{\rm I/III,\pm}(\rho,\phi,z)
   &=& \frac{R J_{m}(k_{m,j}\rho) \exp(im\phi)}
            {\sqrt{2 \int_{0}^{R} d\rho \rho
                    [J_{m}(k_{m,j}\rho)]^{2}}}
       \exp(\pm i \sqrt{ \frac{2m}{\hbar^{2}}E-k_{m,j}^{2}} z).
\end{eqnarray}
The radial wave vectors $k_{m,j}$ that characterizes these states
are solutions of $J_{m}^{\prime}(k_{m,j}R)=0$. This condition of
vanishing radial derivative of the wave function
on the border of the cylinder is imposed in
the entire system (Region II included). It enables the wire to
allow for at least one solution, namely $k_{0,0}$=0, for any value of
the energy $E$.
The way the electronic states are propagated through Region II is
explained with details in Refs \citenum{Mayer1}, \citenum{Mayer2},
\citenum{Mayer70} and \citenum{Mayer100}. In order to
improve the clarity of the results, only axially symmetric states will
be considered.

\subsection{Transmission and band structure for a $V(z) = V_{0}
  \cos(\frac{2\pi}{a}z)$ potential}

The first potential we consider is given by $V(z) = V_{0}
  \cos(\frac{2\pi}{a}z)$, with $V_{0}$= 0.4 eV and $a$=0.434 nm.
These parameters are chosen so that a 0.4 eV-wide band gap
appears at an electron energy of
$\frac{\hbar^{2}}{2m}(\frac{\pi}{a})^2$ = 2 eV.
After calculation of the transfer matrices associated with a
single period $a$ of the potential and using the layer-addition
algorithm presented in Sec. \ref{layer-addition},
it is straightforward to compute how the electronic transmission in
the wire changes as the number of periods increases.

\begin{figure}[h]
 \begin{center}
  \begin{tabular}{c}
   \includegraphics[height=8cm,angle=-90]{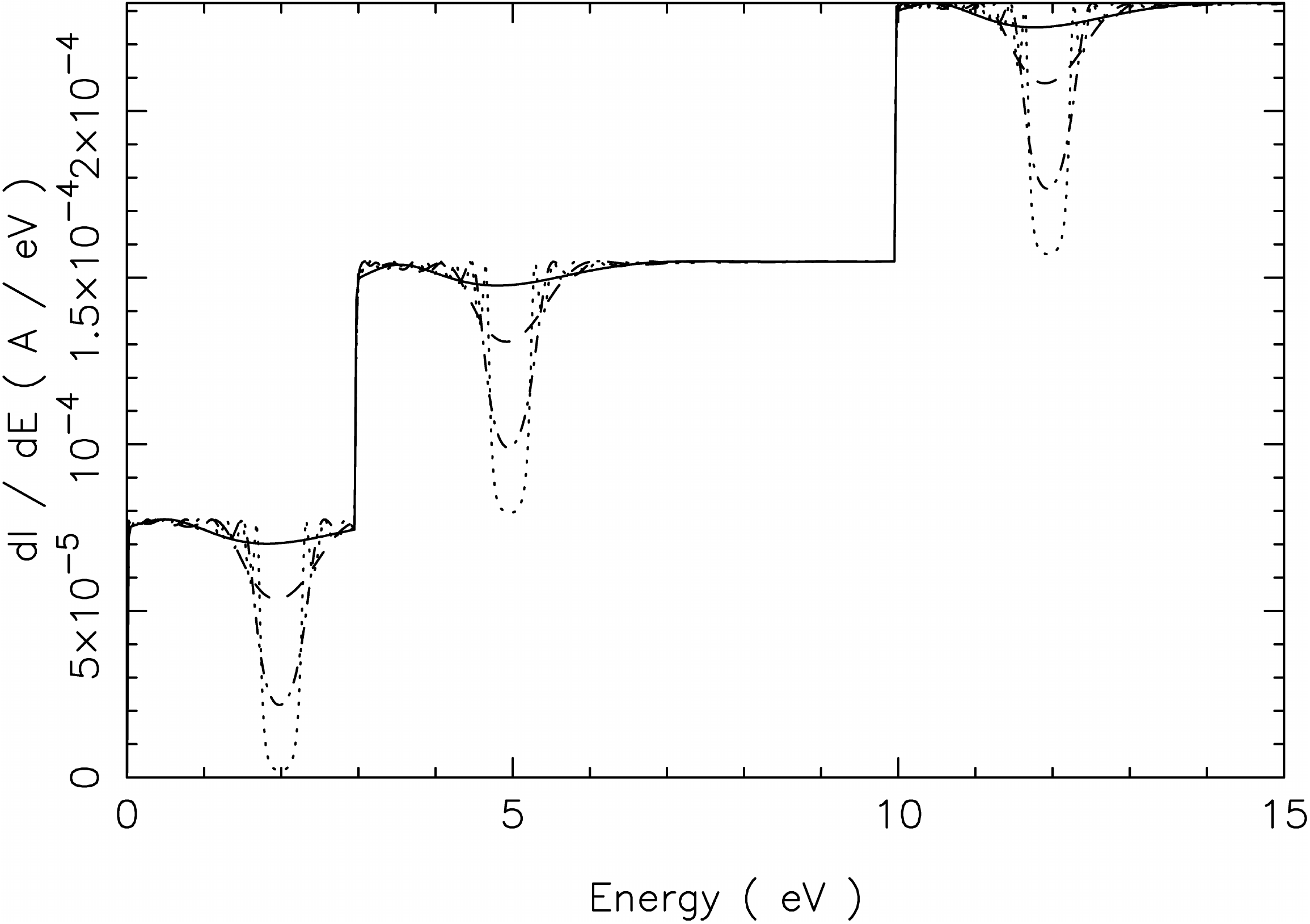}
  \end{tabular}
 \end{center}
 \caption{\label{figure2}Values of $dI/dE$ after 2 (solid), 4 (dotted), 8 (dot-dashed) and 16 (dotted) periods of a
  $V(z) = 0.4 \cos(\frac{2\pi}{a}z)$ eV potential in a cylinder with radius $a$=0.434 nm.}
\end{figure}

We illustrated in Fig. \ref{figure2} the electronic transmission (more precisely
the values of $dI/dE$) for tube lengths corresponding to 2, 4, 8 and
16 periods of the potential and electron energies ranging from 0 to 15 eV.
One can observe the apparition of gaps, which tend to be more
pronounced as the number of periods increases. Besides the gaps, the
transmission tends to its maximal value and exhibits oscillations
that are related to stationary waves in the structure.
Indeed their number and the sharpness of their contribution in the
transmission diagram increase with the number of periods.
Similar observations were made when studying the conduction and
field-emission properties of the semiconducting (10,0) carbon
nanotube.\cite{Mayer27}

These non-zero values of the transmission at energies where a band-gap
exists when the medium is infinite is due to the finite length of the
structures considered here and to the existence of exponentially
decaying solutions in these regions. It is only in truly infinite structures
that these solutions are prohibited because of their exploding
behavior at either $z=+\infty$ or $-\infty$. The existence of decaying
solutions in band-gaps was invoked in Ref. \citenum{Mayer30}
to justify the presence of photon-excited electrons in the gap of a
nanometer-size (10,0) carbon nanotube in a context of field emission.

\begin{figure}[h]
 \begin{center}
  \begin{tabular}{c}
   \includegraphics[height=8cm,angle=-90]{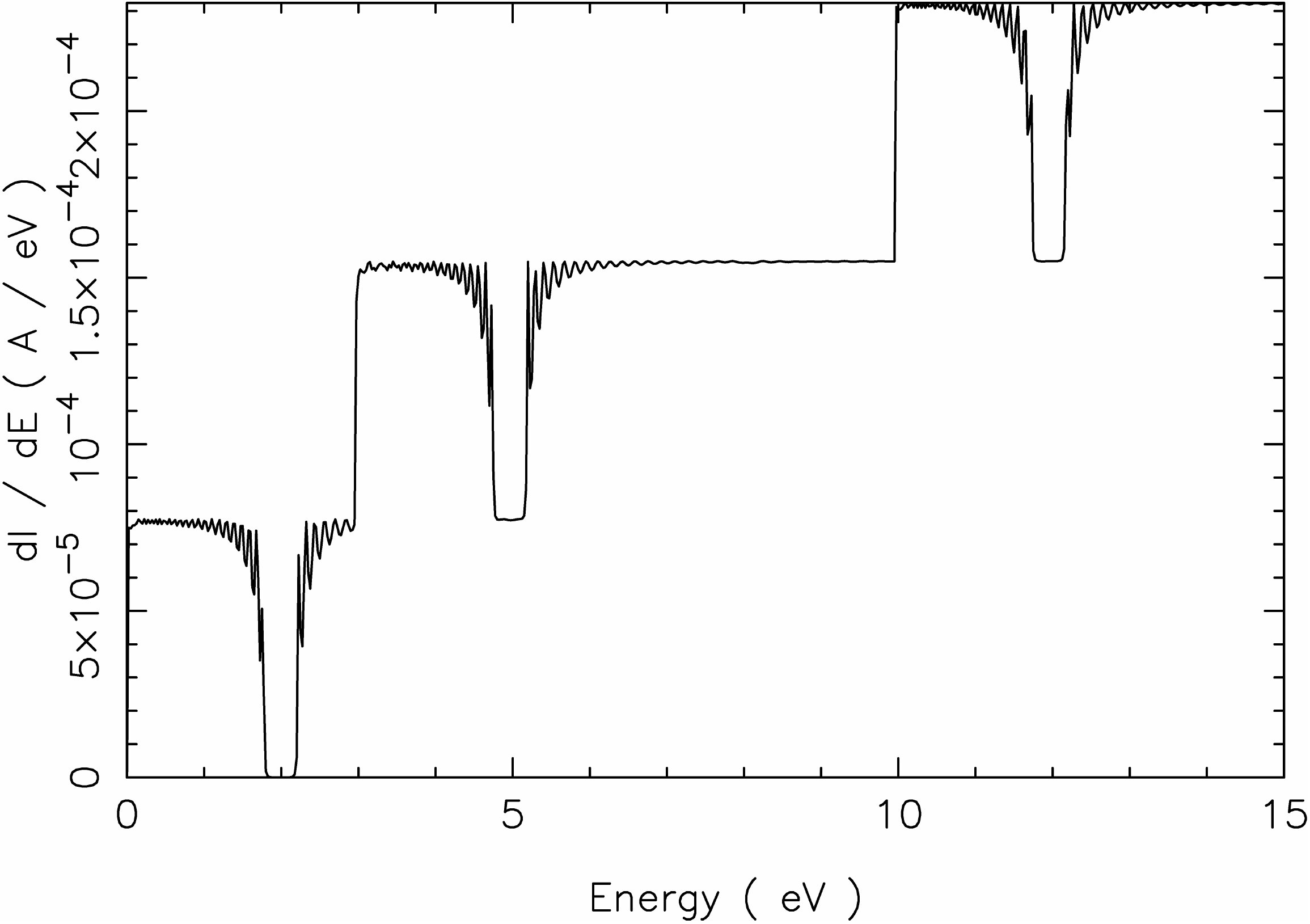} \cr
   \includegraphics[height=8cm,angle=-90]{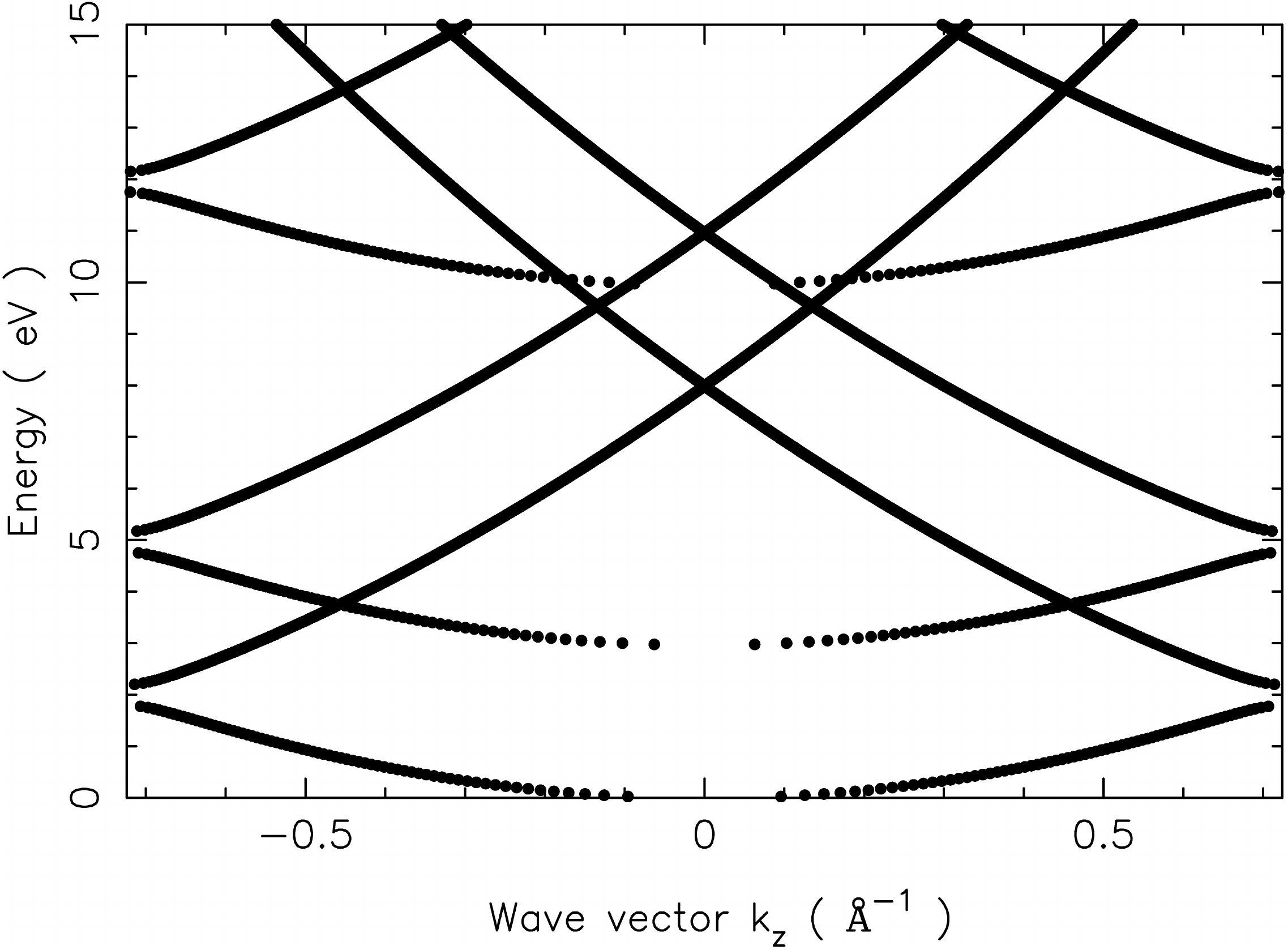} \cr
  \end{tabular}
 \end{center}
 \caption{\label{figure3}Top: values of $dI/dE$ after 32 periods of a
  $V(z) = 0.4 \cos(\frac{2\pi}{a}z)$ eV potential in a cylinder with
  radius $a$=0.434 nm. Bottom: band structure characterizing the
  infinite medium.}
\end{figure}

The values of the transmission after 32 periods of the potential and
the band structure characterizing the infinite medium are
represented in Fig. \ref{figure3}. The gaps in the transmission
diagram are now well pronounced and in agreement with those in the
band structure.
There is a step in the transmission each time the energy is sufficient
to allow for a new state in the radial direction. The occurrence of
these steps coincide with the beginning of new bands in the band
structure.
The height of the steps is given by $2e^{2}/h$
(7.74$\times$10$^{-5}\Omega^{-1}$), which is twice the value of the
conductance quantum since each basis state is representative of
two electrons with opposite spins.
Because of the value of the period $a$, the gaps follow always by 2 eV
the steps in the transmission diagram, which reflects the fact
that the band-gaps are always 2 eV higher in energy than the beginning
of the new bands.

It is interesting to notice that the energy where all transitions or gaps
appear are close to integer values in eV ! In particular, the steps
associated with new solutions appear at 3 and 10 eV. This peculiarity
can be explained by the fact that the solutions of the boundary condition
$J_{0}^{\prime}(k_{0,j}a)=0$ are given in a first approximation
by $k_{0,j}a = (j+1/4) \pi$.\cite{Abramowitz1}
If we remember that $a$ was chosen so that
$\frac{\hbar^{2}}{2m}(\frac{\pi}{a})^2$ = 2 eV, it can easily be shown
that the energy associated with the lateral wave vectors $k_{0,j}$ is
given approximately by $E_{0,j}= (2 j^{2} + j + 1/8)$ eV, which
explains our observations and predicts the position of the next steps.

\subsection{The issue of bound states with a $V(\rho) = V_{0}
  \cos(\frac{2\pi}{T}\rho)$ potential}

We will now address the issue of bound states in the intermediate Region II
and the related question of the number of basis states to consider in
this region when doing a transfer-matrix calculation.
In the boundary Region I and Region III, the number of basis states is fixed by the condition
${\hbar^{2}k_{m,j}^{2}}/{2m} \leq E$. Indeed basis states with higher
$k_{m,j}$ values would be real exponentials in the $z$ direction,
carrying no current and causing only instabilities
(we assume the potential energy to be zero in Region I and Region III).
Since however the potential energy in the intermediate Region II can take
negative values, the condition on $k_{m,j}$ inside Region II must be relaxed to
${\hbar^{2}k_{m,j}^{2}}/{2m} \leq E+\Delta E$ and there is the possibility
for this region to accommodate additional states, which are
exponentially decreasing outside this region but not inside.

This raises an issue on the necessity to consider these {\it bound states}
or not when doing a transfer-matrix calculation. A first technical difficulty arises from the fact the number of states
in Region II is different from that in Region I and Region III. Because of that,
connecting the solutions at $z=0$ and $z=D$ involves the inversion of non-square
matrices.\cite{note} All techniques required to deal efficiently with this point were
however developed in Ref. \citenum{Mayer9}.
Another point is that, according to the literature and for different
formulations of the transfer-matrix methodology,\cite{Bayman1,Wu1}
these bound states are likely to cause numerical instabilities.
Our objective was therefore to create artificially bound states in our system
and study their effect as well as the necessity to consider them or not.

Let us first consider a $V(\rho) = V_{0} \cos(\frac{2\pi}{T}\rho)$ potential,
with $V_{0}$=1 eV.
The length $D$ and radius $R=2T$ of the cylinder are increased to 1 nm.
Because of its $\rho$-dependence - and unlike the previous case -
this potential introduces a coupling between the basis states in Region II,
which is a necessary condition to observe any effect associated with bound states.
Since the potential is independent of $z$, its effect is actually to
redefine the electronic states that propagate independently in the
wire from the original ones (i.e., states characterized by given
values of $m$ and $j$) to combinations of them and one can already
understand the necessity to a have enough basis states to represent
these new states correctly.

\begin{figure}[h]
 \begin{center}
  \begin{tabular}{c}
   \includegraphics[height=8cm,angle=-90]{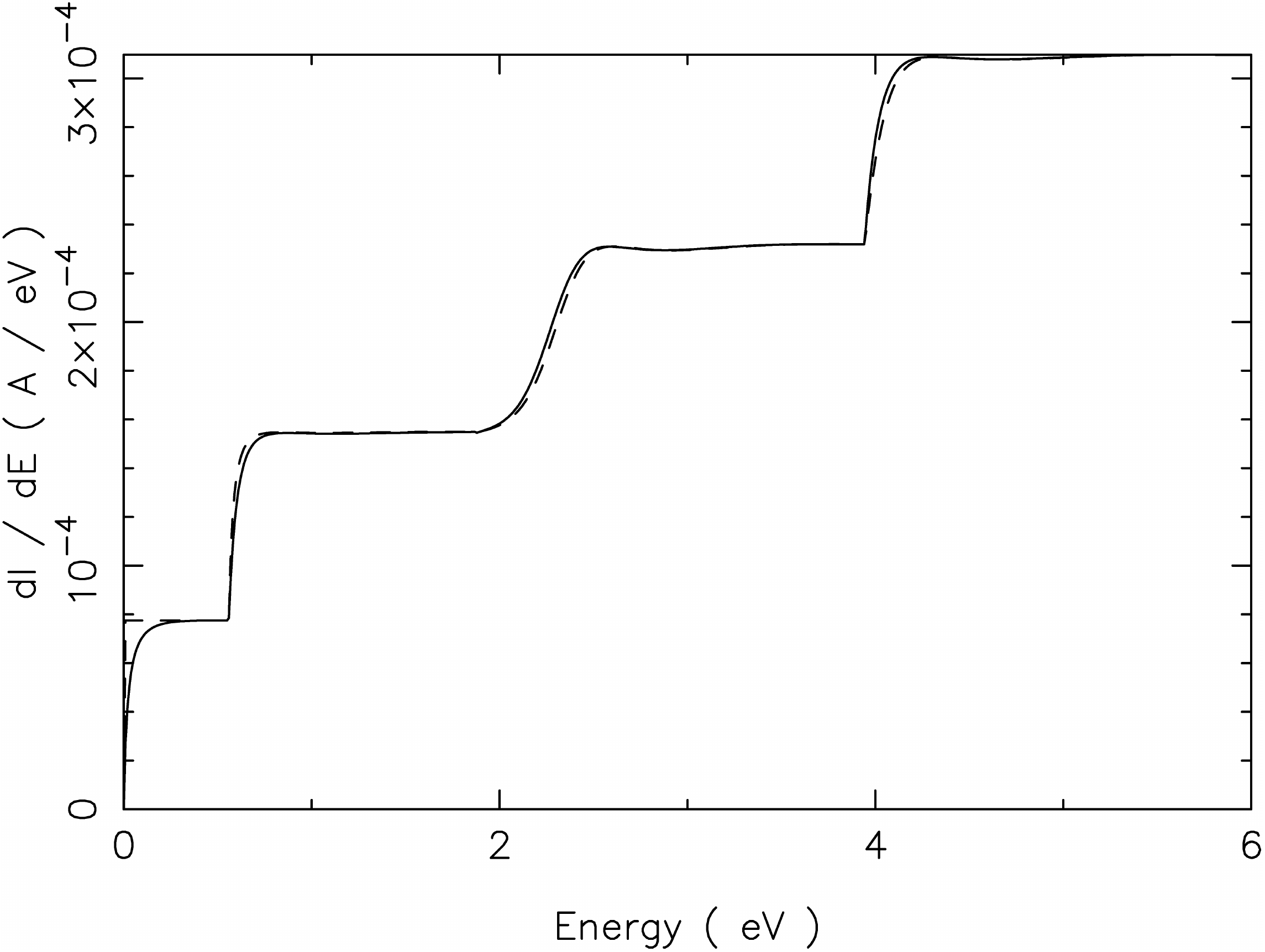} \cr
   \includegraphics[height=8cm,angle=-90]{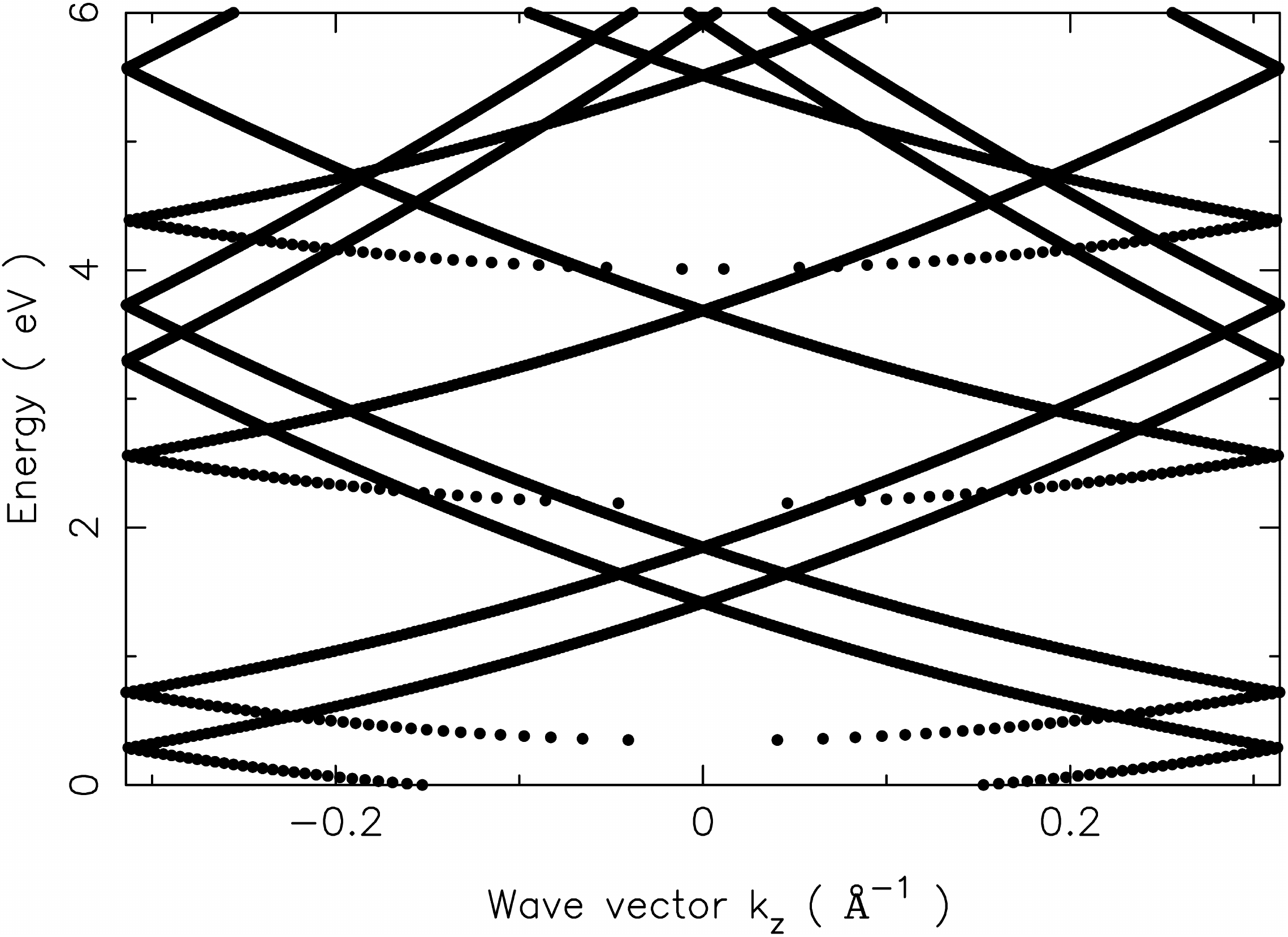} \cr
  \end{tabular}
 \end{center}
 \caption{\label{figure4}Top: values of $dI/dE$ after $D$=1 nm of a $V(\rho) = \cos(\frac{2\pi}{T}\rho)$ eV potential in a
  cylinder with radius $R=2T$=1 nm. The solid curve corresponds to $\Delta E$ = 20 eV and the dashed one to $\Delta E$ = 0.
  Bottom: band structure characterizing the infinite repetition of Region II.}
\end{figure}

We represented in Fig. \ref{figure4} the values of $dI/dE$ obtained at
$z=D$ as well as the band structure characterizing the infinite repetition
of Region II.
These results were obtained by considering $\Delta E$=20 eV, i.e. nine basis
states within Region II while there are only four of them in Region
I and Region III.
These additional states serve essentially to remove unphysical discontinuities
in the band structure, which appear when the degree of completeness of the basis is
poor. The role of $\Delta E$ is identical to the "cut-off energy" in plane-waves
calculations and there is no effect associated with bound states, which are not
present here. We checked that these results keep unchanged when considering
higher values of $\Delta E$ (up to 50 eV). For the purpose of comparison we
represented the values of $dI/dE$ obtained with $\Delta E$=0 (showing that the
currents are less sensitive to the completeness of the basis than the band
structures).

\begin{figure}[h]
 \begin{center}
  \begin{tabular}{c}
   \includegraphics[height=8cm,angle=-90]{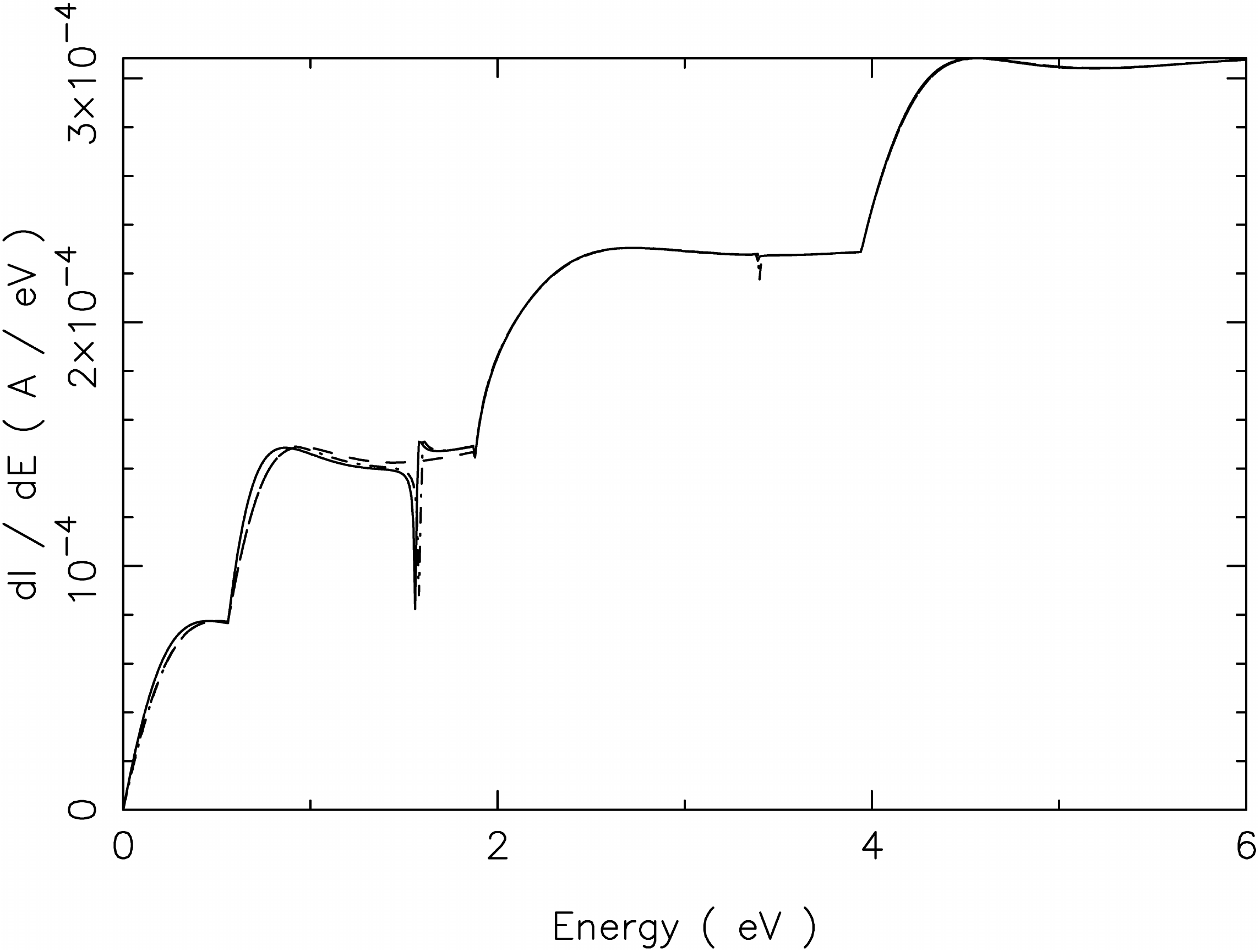} \cr
   \includegraphics[height=8cm,angle=-90]{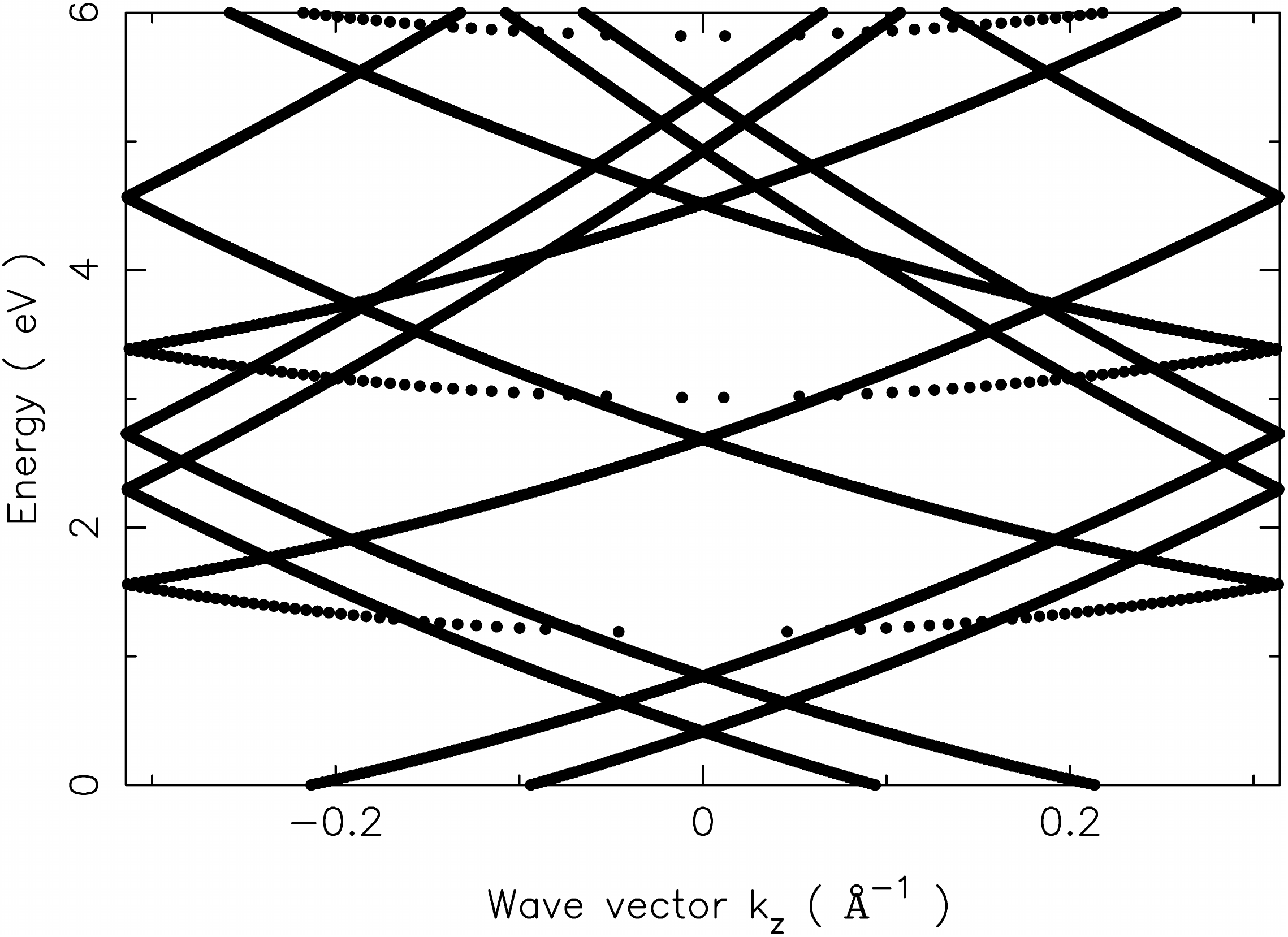} \cr
  \end{tabular}
 \end{center}
 \caption{\label{figure5}Top: values of $dI/dE$ after $D$=1 nm of a $V(\rho) = \cos(\frac{2\pi}{T}\rho)$-1 eV potential in a
  cylinder with radius $R=2T$=1 nm. The solid curve corresponds to $\Delta E$ = 20 eV, the dashed one to $\Delta E$ = 0 and the
  dot-dashed one to $\Delta E$=1 eV. Bottom: band structure characterizing the infinite repetition of Region II.}
\end{figure}

Let us now consider the $V(\rho) = V_{0} \cos(\frac{2\pi}{T}\rho) - 1$ eV potential.
We represented in Fig. \ref{figure5} the corresponding values of $dI/dE$ as
well as the band structure that would characterize Region II if
repeated periodically. As expected, the band structure is shifted down by 1 eV.
This means that the two bands that stood between 0 and 1 eV in Fig. \ref{figure4}
now give rise to discrete energy levels, characterizing bound states. The
position of these energy levels is given by the intersection of the former bands with
the limits $\pm \pi/D$ of the first Brillouin zone (namely at -0.71 and -0.28 eV),
since the length $D$ of Region II is then an integer multiple of half the electronic
wave length in the $z$ direction. For the same reason, quasi-bound states in the
continuum part of the spectrum ($E \geq 0$) will exist each time the bands
of Fig. \ref{figure5} meet the limits $\pm \pi/D$ of the first Brillouin zone.

Despite the fact these bound states only exist in Region II, they have an
impact on the propagative solutions in the $E \geq 0$ range.
As observed in previous work,\cite{Bayman1,Price2,Price3,Price1} this impact is essentially limited
to localized resonances in the $dI/dE$ values, at energies where the
interaction between propagative states and (quasi-)bound states is stronger.
Indeed the two resonances in Fig. \ref{figure5} appear at energies where bands
meet the border of the first Brillouin zone for the first time
(the electronic wave length in the $z$ direction is then identical to
that of the bound states, which enhances the interactions).

The results presented here were obtained by taking $\Delta E$ = 20 eV, as required
for the completeness of the basis. In particular it is necessary to consider
the two bound states (which are part of the solution within Region II).
Neglecting them by taking $\Delta E$=0 has indeed a strong impact on both
the band structure (all bands are truncated at their beginning on the first eV) and the
$dI/dE$ values (the resonances disappear). As illustrated in Fig. \ref{figure5},
taking $\Delta E$ = 1 eV is sufficient to include the bound states and therefore
reproduce the resonances and complete the bands. The additional states
introduced by taking higher values of $\Delta E$ only improve the completeness of
the basis and serve essentially to remove unphysical discontinuities in the bands.

The relation between resonances in the transmission currents and quasi-bound
states in the system was well described by Price,\cite{Price2,Price3,Price1}
who actually relates them to poles of the $S$ matrix (whose elements are
considered as functions of the energy). The present simulations show that
these effects are addressed properly, provided additional basis states are
considered in the intermediate Region II (through $\Delta E > 0$).
The "interior states" do not need to be computed explicitly, nor
treated differently from "open states".\cite{Bayman1,Wu1} The specificity
of our approach is to use non-square transfer matrices\cite{Mayer9}
to prevent instabilities when making the connection between the
different regions.

\section{Conclusions}

This paper was a pedagogical presentation of the transfer-matrix technique,
with an extension to extract the band structure of periodic
materials from the $T$ or $S$ matrices associated with a single unit.
Because of the transfer-matrix formulation of the scattering problem,
the band structure is projected on the $k_{z}$ axis, which is often
appropriate in situations where this technique can be applied.

We provided calculations of the transmission and
band structure of electrons confined in a cylindrical wire and subject
to cosine potentials. We observed how fast the transmission diagram
exhibits characteristics predicted by the band structure
(namely gaps and steps associated with the opening of new bands), while
keeping features associated with their finite length.
Comparisons could be made with results obtained for a semiconducting
(10,0) carbon nanotube, confirming and providing an insight on processes observed
in complex structures.

The issue of bound states was considered. Although they exist only in the
intermediate region, they need to be included in the representation
because of their impact
on propagative solutions (as localized resonances in the transmission) and
to avoid unphysical truncations of the band structure. Considering additional
states essentially improves the completeness of the representation and
removes discontinuities in the bands.
The connection between the intermediate region
(which may contain "interior states") and the boundary regions
(which contain only propagative states) is achieved using
non-square transfer matrices, making the technique perfectly stable.

\acknowledgments A.M. is funded by the Fund for Scientific Research
(F.R.S.-FNRS) of Belgium. He is member of NaXys, Namur Institute for
Complex Systems, University of Namur, Belgium.
The author acknowledges the use of the Namur Scientific
Computing Facility and the Belgian State Interuniversity Research
Program on {\it Quantum size effects in nanostructured materials}
(PAI/IUAP P5/01). P.H. Cutler, N.M. Miskovsky and Ph. Lambin
are acknowledged for useful discussions.

\appendix
\section{\label{Appendix_A}Derivation of band structures from the ${\bf S}^{\pm\pm}$ matrices}

An alternative method for extracting a band structure from the
${\bf S}^{\pm\pm}$ matrices that characterize the basic unit
of a periodic system was derived by Pendry.\cite{Pendry1} This method
is adapted in this Appendix to our formulation of the transfer-matrix technique.

The idea consists in considering the infinite, periodic repetition of
a basic unit of length $a$. We will consider that the interfaces between
adjacent units are situated at $z=n.a$, with $n$ an integer.

A wave function $\Psi$ can be developed at these interfaces as
\begin{eqnarray}
 \Psi(z=n.a) = \sum_j c^+_j[n]\ \Psi^+_j(z=n.a) + \sum_j c^-_j[n]\ \Psi^-_j(z=n.a),
\end{eqnarray}
where $\Psi^\pm_j(z=n.a)$ refers to the basis states used at $z=n.a$ for the expansion
of the wave function. $c^\pm_j[n]$ refers to the coefficients of this expansion
at $z=n.a$.

From the physical interpretation of the ${\bf S}^{\pm\pm}$ matrices that describe
the basic unit cell in the interval $z\in[0,a]$, we can write that
\begin{eqnarray}
 {\bf c}^+[1] &=& {\bf S}^{++}\ {\bf c}^+[0] + {\bf S}^{+-}\ {\bf c}^-[1], \\
 {\bf c}^-[0] &=& {\bf S}^{-+}\ {\bf c}^+[0] + {\bf S}^{--}\ {\bf c}^-[1],
\end{eqnarray}
where ${\bf c}^{\pm}$ refers to vectors that contain the coefficients $c_j^\pm$.
We can reorganize these relations and write them like
\begin{eqnarray}
 \begin{pmatrix}
   {\bf S}^{++} & {\bf 0} \cr
  -{\bf S}^{-+} & {\bf I} \cr
 \end{pmatrix}
 \begin{pmatrix}
  {\bf c}^+[0] \cr
  {\bf c}^-[0] \cr
 \end{pmatrix}
=
 \begin{pmatrix}
  {\bf I} & -{\bf S}^{+-} \cr
  {\bf 0} &  {\bf S}^{--} \cr
 \end{pmatrix}
 \begin{pmatrix}
  {\bf c}^+[1] \cr
  {\bf c}^-[1] \cr
 \end{pmatrix}.\label{Appendix_relation}
\end{eqnarray}

We are looking for Bloch-state solutions that satisfy $\Psi(z=a)=e^{ik_z a}\ \Psi(z=0)$.

In the context of this paper, we can write that
\begin{eqnarray}
 \Psi(z=0)
&=& \sum_j c^+_j[0]\ \Psi^{{\rm I},+}_j(z=0) + \sum_j c^-_j[0]\ \Psi^{{\rm I},-}_j(z=0) \nonumber \\
&=&
( \Psi_{j}^{\rm I,+} \ldots \Psi_{j}^{\rm I,-} )|_{z=0}
 \begin{pmatrix}
  {\bf c}^+[0] \cr
  {\bf c}^-[0] \cr
 \end{pmatrix}\label{Appendix_Psi_z=0}
\end{eqnarray}
and
\begin{eqnarray}
 \Psi(z=a)
&=& \sum_j c^+_j[1]\ \Psi^{{\rm III},+}_j(z=a) + \sum_j c^-_j[1]\ \Psi^{{\rm III},-}_j(z=a) \nonumber \\
&=&
( \Psi_{j}^{\rm III,+} \ldots \Psi_{j}^{\rm III,-} )|_{z=a}
 \begin{pmatrix}
  {\bf c}^+[1] \cr
  {\bf c}^-[1] \cr
 \end{pmatrix}.
\end{eqnarray}

We have also that
\begin{eqnarray}
 ( \Psi_{j}^{\rm III,+} \ldots \Psi_{j}^{\rm III,-} )|_{z=a}
 =
 ( \Psi_{j}^{\rm I,+} \ldots \Psi_{j}^{\rm I,-} )|_{z=0}\ {\bf diag}[e^{i k_{z,j} a}, \ldots, e^{-i k_{z,j}a}].\label{Appendix_basis_transfo}
\end{eqnarray}

By using Eqs \ref{Appendix_relation} and \ref{Appendix_basis_transfo}, we can develop $\Psi(z=a)$ as
\begin{eqnarray}
 \Psi(z=a)
=
( \Psi_{j}^{\rm I,+} \ldots \Psi_{j}^{\rm I,-} )|_{z=0}\ {\bf diag}[e^{i k_{z,j} a}, \ldots, e^{-i k_{z,j}a}]
 \begin{pmatrix}
  {\bf I} & -{\bf S}^{+-} \cr
  {\bf 0} &  {\bf S}^{--} \cr
 \end{pmatrix}^{-1}
 \begin{pmatrix}
   {\bf S}^{++} & {\bf 0} \cr
  -{\bf S}^{-+} & {\bf I} \cr
 \end{pmatrix}
 \begin{pmatrix}
  {\bf c}^+[0] \cr
  {\bf c}^-[0] \cr
 \end{pmatrix}.\nonumber \label{Appendix_Psi_z=a}\\
\end{eqnarray}

By using Eqs. \ref{Appendix_Psi_z=0} and \ref{Appendix_Psi_z=a} in the equation
$\Psi(z=a)=e^{ik_z a}\ \Psi(z=0)$, we finally obtain that
\begin{eqnarray}
{\bf diag}[e^{i k_{z,j} a}, \ldots, e^{-i k_{z,j}a}]
 \begin{pmatrix}
  {\bf I} & -{\bf S}^{+-} \cr
  {\bf 0} &  {\bf S}^{--} \cr
 \end{pmatrix}^{-1}
 \begin{pmatrix}
   {\bf S}^{++} & {\bf 0} \cr
  -{\bf S}^{-+} & {\bf I} \cr
 \end{pmatrix}
 \begin{pmatrix}
  {\bf c}^+[0] \cr
  {\bf c}^-[0] \cr
 \end{pmatrix}
=
 e^{ik_z a}\
 \begin{pmatrix}
  {\bf c}^+[0] \cr
  {\bf c}^-[0] \cr
 \end{pmatrix}
\end{eqnarray}
or equivalently the generalized eigenvalue system
\begin{eqnarray}
 \begin{pmatrix}
   {\bf S}^{++} & {\bf 0} \cr
  -{\bf S}^{-+} & {\bf I} \cr
 \end{pmatrix}
 {\bf x}
=
 \lambda\
 \begin{pmatrix}
  {\bf I} & -{\bf S}^{+-} \cr
  {\bf 0} &  {\bf S}^{--} \cr
 \end{pmatrix}
 {\bf diag}[e^{-i k_{z,j} a}, \ldots, e^{i k_{z,j}a}]\
 {\bf x},\label{Appendix_bandstructure_system}
\end{eqnarray}
where the generalized eigenvalue $\lambda$ will provide the factor $e^{ik_z a}$
and the corresponding eigenvector ${\bf x}$ actually provides the coefficients $c^\pm_j[0]$
of the Bloch-state solution at $z=0$.

Eq. \ref{Appendix_bandstructure_system} was used by Mayer in Ref. \citenum{Mayer_2004} to compute the
band structure of carbon nanotubes. The approach described in this Appendix turns out to be numerically
more stable than Eq. \ref{bandstructure_system} of the text, which relies on the ${\bf T}^{\pm\pm}$ matrices.


\begin{thebibliography}{99}
\bibitem{note} Indeed a reference potential of $-\Delta E$ is assumed between
 adjacent layers to maintain stability, so the calculations within $[0,D]$ only
 involve square transfer matrices. It is only when connecting the solutions
 with those in Region I and Region III, where the reference potential is zero, that
 non-square transfer matrices are needed.
\end{thebibliography}
\end{document}